\documentclass[12pt,twoside,fleqn]{article}
\usepackage{epsfig}
\usepackage{psfig}
\def\lsim{\raise0.3ex\hbox{$<$\kern-0.75em\raise-1.1ex\hbox{$\sim$}}}
\def\gsim{\raise0.3ex\hbox{$>$\kern-0.75em\raise-1.1ex\hbox{$\sim$}}}
\makeatletter
\@addtoreset{equation}{section}
\makeatother

\newcommand{\beqn} {\begin{equation}}
\newcommand{\eqn} {\end{equation}}

\newcommand{\slsh}[1] {#1\kern-.43em/}
\newcommand{\real}{{\sf I}\kern-.12em{\sf R}}
\newcommand{\comp}{{\sf I}\kern-.48em{\sf C}}
\newcommand{\nin} {\in\kern-.6em/}

\def\ie{{\sl i.e.\/}}

\def\MEF{m_{\rm eff}}\def\mef{\ifmmode\MEF\else$\MEF$\fi}

\def\SM{s_{\mu}}\def\xm{\ifmmode\SM\else$\SM$\fi}
\def\sm{s_\mu}
\def\smb{s_\mu^3 }
\def\smc{s_\mu^5 }
\begin{document}
\thispagestyle{empty}
%
 \mbox{} \hfill CERN-TH/95-253 \\
 \mbox{} \hfill BI-TP 95/33\\
 \mbox{} \hfill arch-ive/9510031\\
\begin{center}
{{\large \bf Improved Actions for QCD Thermodynamics on the Lattice}
 } \\
\vspace*{1.0cm}
{\large B. Beinlich$^{1}$, F. Karsch$^{1,2}$ and E. Laermann$^1$} \\
\vspace*{1.0cm}
{\normalsize
$\mbox{}$ {$^1$Fakult\"at f\"ur Physik, Universit\"at Bielefeld,
D-33615 Bielefeld, Germany}\\
$\mbox{}$ {$^2$Theory Division, CERN, CH-1211, Geneva, Switzerland}
}
\end{center}
\vspace*{1.0cm}
\centerline{\large ABSTRACT}

\baselineskip 20pt

\noindent
Finite cut-off effects strongly influence the thermodynamics of
lattice regularized QCD at high temperature in the standard Wilson
formulation. We analyze the reduction of finite cut-off effects in
formulations of the thermodynamics of $SU(N)$ gauge theories
with three different $O(a^2)$ and $O(a^4)$ improved actions.
We calculate the energy density and pressure on finite lattices in leading
order weak coupling perturbation theory ($T\rightarrow \infty$) and perform
Monte Carlo simulations with improved $SU(3)$ actions at non-zero $g^2$.
Already on lattices with temporal extent $N_\tau=4$ we find a strong reduction
of finite cut-off effects in the high temperature limit, which persists also
down to temperatures a few times the deconfinement transition temperature.
\vskip 20pt
\vfill
\baselineskip 15pt
\noindent
CERN-TH/95-253 \\
BI-TP 95/33\\
October 1995\\
\eject
\baselineskip 15pt

\section{Introduction}

With increasing accuracy of lattice calculations it also becomes increasingly
important to control and, as far as possible, eliminate the systematic errors
introduced through the finite lattice cut-off in numerical calculations.
Various procedures to achieve this goal without increasing the necessary
computational effort drastically have been suggested in the past, including
the early {\it Symanzik improvement} programme \cite{Sym83}, renormalization
group improved actions \cite{Iwa84} as well as the
recently constructed perfect actions \cite{Has94,DeG95}. The resulting actions
are not designed to lead to a better infrared behaviour of lattice regularized
field theories.
It thus may not have come as a surprise, that the application of these actions
in lattice calculations, which aimed at a determination of long-distance
properties of asymptotically free field theories, did show little advantages
over the standard one-plaquette action originally proposed by
Wilson \cite{Wil74}. In particular, this is the case for studies of the
finite temperature deconfinement transition in $SU(N)$ gauge theories with
Symanzik-improved actions \cite{Cel94a,Cel94b}.
On the other hand the improved actions are expected to lead to a better
ultraviolet behaviour of the theory. This has, for instance, been observed
in studies of topological properties \cite{vanB} or the short distance
part of the heavy quark potential \cite{DeG95}. In these cases the unwanted
short distance lattice artifacts could successfully be eliminated.

The high temperature behaviour of QCD is close to that of an ideal gas.
Bulk thermodynamic quantities are therefore dominated by contributions from
large momenta. These, however, are most strongly influenced by finite cut-off
effects. One thus may expect that improved actions will be particularly
useful for the analysis of bulk thermodynamic quantities.
In fact, it is well known that in the standard Wilson formulation
of lattice QCD the cut-off effects lead to strong modifications of the
high temperature limit of quantities like, e.g. the energy density and
pressure \cite{Eng82,Hel85}. These cut-off effects are $O((aT)^2)$ in the
pure gauge sector.
Calculations on lattices with temporal extent $N_\tau =1/aT$ have recently
been performed for the $SU(2)$ \cite{Eng95} and $SU(3)$ \cite{Boyd95}
gauge theories. They clearly show the influence of a finite ultraviolet
cut-off on the behaviour of bulk thermodynamic quantities and their
extrapolation to the continuum limit. In the standard Wilson formulation
\cite{Wil74} lattices with temporal extent $N_\tau \gsim 8$ are needed in
order to reduce deviations of, eg. $\epsilon /T^4$,
from the continuum extrapolation below a few percent. The interesting bulk
thermodynamic quantities like energy density and pressure all have dimension
four. In lattice calculations they are determined from operators,
whose expectation values are proportional to $N_\tau^{-4} = (aT)^4$. It
thus rapidly becomes difficult to calculate these operators with
reasonable statistical significance. In fact, the effort to calculate them
with a given accuracy on lattices of size $N_\sigma^3 \times N_\tau$,
increases at least like $N_\tau^{11}$, if one also keeps the physical size
of the thermodynamic system constant ($N_\sigma/N_\tau =$~const.).
It therefore should be clear that a huge
improvement is already achieved, if one can perform calculations on
lattices with, say $N_\tau =4$, with cut-off distortions similar in magnitude
to what one reaches in calculations with the Wilson action on lattices with
twice that temporal extent.

It is the purpose of this paper to quantify the ultraviolet cut-off effects
that can be expected to be present in calculations of thermodynamic
quantities with improved actions and to examine the relevance of improved
actions for the calculation of the equation of state in QCD at high
temperature. We will discuss
various improved actions and calculate finite cut-off corrections to
the high temperature limit of the energy density in leading order perturbation
theory. We also will present some numerical results for the pressure of a
$SU(3)$ gluon gas calculated with an $O((aT)^2)$ improved action. We are
going to discuss tree level improved actions ($g^2\equiv 0$).
As will become clear in the following this leads already to a significant
improvement of the high temperature behaviour of bulk thermodynamic
quantities in $SU(N)$ gauge theories. There is,
however, no fundamental problem which would prohibit the extension of
the present considerations to $O(g^2)$ improved actions and we will
present evidence that this is, in fact, needed for temperatures close
to the deconfinement transition temperature.

We will present in the next section two extensions of the
standard one-plaquette Wilson action leading to an $O((aT)^2)$ improvement
of thermodynamic observables as well as a specific choice of an action
that leads to an improvement at $O((aT)^4)$. In section 3 we will discuss the
perturbative high temperature limit of these  improved actions. A first
exploratory numerical analysis of these actions and a comparison with
the standard one-plaquette Wilson action  is presented in section 4.
In section 5 we give our conclusions.
Various details of our perturbative calculations are given in an Appendix.

\section{Improved SU(N) Actions}

In order to eliminate the $O(a^2)$ and higher order corrections to the
lattice version of the Euclidean action one can add appropriately chosen
larger loops to the basic 4-link plaquette term appearing in the
standard one-plaquette Wilson action. We will, in particular, discuss
here simple extensions of the one-plaquette action, which only involve
larger planar loops, {\it i.e.} we will consider the generalized Wilson
actions,

\begin{eqnarray}
S^{I} & = &   \sum_{x, \nu > \mu} \sum_{l=1}^\infty  \sum_{k=1}^l
\; a_{k,l}\; W^{k,l}_{\mu, \nu}(x) \nonumber \\
& \equiv &  \sum_{x, \nu > \mu} S^{I}_{\mu, \nu} (x)
{}~~,
\label{actionnn}
\end{eqnarray}
where $W^{k,l}_{\mu, \nu}$ denotes a symmetrized combination of $k\times l$
Wilson loops in the $(\mu, \nu)$-plane of the lattice,
\beqn
W^{k,l}_{\mu, \nu} (x) = 1- {1 \over 2N} \biggl( {\rm Re~Tr} L^{(k)}_{x,\mu}
L^{(l)}_{x+k\hat\mu,\nu} L^{(k)+}_{x+l\hat\nu,\mu} L^{(l)+}_{x,\nu}
+ (k \leftrightarrow l) \biggr)~~.
\label{wloop}
\eqn
Here we have introduced the short hand notation for {\it long} links,
$ L^{(k)}_{x,\mu} =\prod_{j=0}^{k-1} U_{x+j\hat\mu,\mu}$ and
$x=(n_1,n_2,n_3,n_4)$ denotes the sites on an
asymmetric lattice of size $N_\sigma^3 \times N_\tau$.

With a suitable choice of the coefficients $a_{k,l}$ it can be achieved that
the generalized Wilson actions reproduce the continuum Euclidean Yang-Mills
Lagrangian,
${\cal L} = -{1\over 2} F_{\mu, \nu}  F_{\mu, \nu}$,  up to some order
$O(a^{2n})$ \cite{Sym83}. The standard one-plaquette Wilson action,
$S^{(1,1)}$, with $a_{1,1}=1$ and
$a_{k,l} =0$ for all $(k,l)\ne (1,1)$ receives $O(a^2)$ corrections in the
naive continuum limit. Expanding
the link variables, $U_{x,\mu} = \exp(igaA_\mu (x))$, in powers of $a$ one
finds
\begin{eqnarray}
S_{\mu,\nu}^{(1,1)} (x) &=& 1- {1\over N} {\rm Re~Tr}
U_{x,\mu}
U_{x+\hat\mu,\nu} U^{+}_{x+\hat\nu,\mu} U^{+}_{x,\nu}
\nonumber \\
&=& -{1\over 2N} g^2 a^4 \biggl( F_{\mu,\nu}  F_{\mu,\nu} + {1\over 12}
a^2  F_{\mu,\nu}(\partial^2_\mu +\partial^2_\nu)  F_{\mu,\nu}
\nonumber \\
& &\qquad\qquad +O(a^4) \biggr) ~~.
\label{expansion}
\end{eqnarray}
By adding
an additional Wilson loop to the action one can achieve that corrections
start only at $O(a^4)$. In particular we will consider here actions
obtained by adding a planar 6-link and 8-link loop, respectively. The
non-vanishing coefficients in these cases are,

\begin{eqnarray}
I\equiv (1,2) :\quad & & a_{1,1}= {5 \over 3}  \quad , \quad
a_{1,2}= -{1 \over 6} \nonumber \\
I\equiv (2,2) :\quad & & a_{1,1}= {4 \over 3} \quad , \quad
a_{2,2} = -{1 \over 48}
\label{actions}
\end{eqnarray}
The action $S^{(1,2)}$ is a specific choice of the 6-link improved
actions originally proposed by Symanzik \cite{Sym83}.
The action $S^{(2,2)}$ has recently also been discussed in the context of
NRQCD calculations \cite{Lep92,Mor93}. Its generalization to larger quadratic
loops gives a straightforward
procedure to eliminate also higher order cut-off effects. Already in
$O(a^4)$ there exist two independent operators of dimension eight, which
contribute to the finite cut-off effects. In general, one thus needs
three independent loops of length six and eight in order to eliminate
all cut-off effects proportional to $a^2$ and $a^4$, respectively.  In an
action constructed only from quadratic loops the different operators
contribute, however, with the same relative weight. In that case it thus
is sufficient to add only two additional, quadratic loops to obtain
an $O(a^4)$ improved action. We will consider here the case where we
add  $2\times 2$ and $3\times 3$ Wilson loops to the plaquette
term\footnote{This procedure can easily be extended
to higher orders. The resulting actions are, however, non-local in the sense
that the magnitude of the coefficients of a $k\times k$ loop asymptotically
only decreases powerlike, $|a_{k,k}| \sim k^{-4}$.}.
In this case corrections to the continuum limit start  only at $O(a^6)$.
The non-vanishing coefficients for the action $S^{(3,3)}$ are given by,
\begin{eqnarray}
I\equiv (3,3) :\quad & & a_{1,1}= {3 \over 2} \quad , \quad
a_{2,2} = -{3 \over 80}
\quad , \quad a_{3,3} = {1 \over 810}
\label{action3}
\end{eqnarray}

In the following we will discuss the behaviour of thermodynamic
quantities in the high temperature limit, which are obtained from
the partition function,
\beqn
Z= \int \prod_{x,\mu} {\rm d} U_{x,\mu}~{\displaystyle e}^{-
{\displaystyle \beta S^{I}}} ~~,
\label{partition}
\eqn
defined on lattices of size $N_\sigma^3\times N_\tau$. Here $\beta=2N/g^2$
is the bare gauge coupling.
Thermodynamic quantities like the energy
density or the pressure are obtained as derivatives of the logarithm of
the partition function with respect to
the temperature, $T=1/N_\tau a$, and the volume, $V=(N_\sigma a)^3$. They are
thus represented by expectation
values of certain parts of the action, {\it i.e.} the operators used
to evaluate thermodynamic quantities are improved to the same order as the
action. In the following we will show explicitly that the $O(a^n)$ improvement
of the action will lead to an $O((aT)^n)$ improvement of thermodynamic
observables.

\section{Perturbative High Temperature Limit}

\subsection{The generalized SU(N) Wilson Actions}

A suitable quantity for the study of the influence of finite cut-off
effects on the thermodynamics of $SU(N)$ gauge theories is
the leading high temperature behaviour of the energy density or equivalently
the pressure. In the limit
$T\rightarrow \infty$ the energy density approaches that of a massless,
non-interacting gluon gas,
$\epsilon /T^4\rightarrow \epsilon_{\rm SB}/T^4 = (N^2-1) \pi^2 / 15$.
Corrections to this are of $O(g^2)$.
Numerical investigations of the high temperature phase of $SU(N)$ gauge
theories show that the energy density rapidly approaches this high temperature
limit. More precisely, on lattices with finite temporal extent
$N_\tau$, it approaches a limiting value,
which differs from $\epsilon_{\rm SB}/T^4$ due
to ultraviolet cut-off effects. At temperatures a few times the
critical temperature deviations from this limiting value are generally of the
same order of magnitude as the cut-off distortion effects.
In order to quantify the approach to the high temperature ideal gas limit
it thus is important to reach a good approximation of this limit already on
finite lattices. The cut-off effects due to finite values of $N_\tau$ can be
studied in leading order perturbation theory. In fact, in the case of the
one-plaquette Wilson action this is known for quite some
time \cite{Eng82,Hel85}.

For the analysis of the high temperature ideal gas limit it
is sufficient to evaluate perturbatively the difference
between expectation values of space- and timelike parts of the action,
\begin{eqnarray}
{\epsilon \over T^4}\equiv 3{p \over T^4}
&=&{6N \over g^2} N_\tau^4 \biggl( \langle
S^{I}_{\sigma}\rangle
-  \langle S^{I}_{\tau}\rangle \biggr) +O(g^2) \nonumber \\
&=& {3\over 2} (N^2 - 1) N_\tau^4 \biggl( S^{I,(2)}_{\sigma} -
S^{I,(2)}_{\tau} \biggr) +O(g^2)~~,
\label{energy0}
\end{eqnarray}
where $S^{I,(2)}_{\sigma}$ and $S^{I,(2)}_{\tau}$ denote the
$O(g^2)$ expansion coefficients of the expectation values of the
space- and timelike parts of the action,

\begin{eqnarray}
\langle  S^{I}_{\sigma} \rangle & \equiv&  {1 \over 3 N_\sigma^3 N_\tau}
\langle
\sum_{x,\mu<\nu<4} S^{I}_{\mu,\nu} (x) \rangle =
g^2 {N^2 - 1 \over 4N}  S^{I,(2)}_{\sigma} + O(g^4) \nonumber \\
\langle  S^{I}_{\tau} \rangle & \equiv&  {1 \over 3 N_\sigma^3 N_\tau} \langle
\sum_{x,\mu=1,2,3} S^{I}_{\mu,4} (x) \rangle =
g^2 {N^2 - 1 \over 4N}  S^{I,(2)}_{\tau} + O(g^4)
{}~~.
\label{weak}
\end{eqnarray}
All other terms appearing in the definition of the energy density
are $O(g^2)$ and thus do not contribute in leading order perturbation
theory \cite{Karsch}.

In the following we will discuss the perturbative calculation of the
expansion coefficients $ S^{I,(2)}_{\sigma}$ and $ S^{I,(2)}_{\tau}$ on finite
lattices. Thereby
we will also review some of the results obtained in perturbative calculations
with the Wilson one-plaquette action on finite, asymmetric lattices
\cite{Eng82,Hel85}. Our notation closely follows \cite{Hel85}. Perturbative
results for the Symanzik action, $S^{(1,2)}$, on infinite lattices have also
been obtained in \cite{Weisz}.

The improved Wilson actions are defined in terms of the link
variables\footnote{In the following the cut-off $a$ has been set to unity.},
$U_{x,\mu} = \exp{(ig A_\mu (x))}$. They can be expanded in
powers of $g^2$ and represented in momentum space.
Sums in Fourier space we denote by
\beqn
\int_p = {1 \over N_\sigma^3 N_\tau} \sum_{(p \ne 0)} \quad ,
\label{fsum}
\eqn
where $p_\mu = (2\pi /N_\sigma) n_\mu$, $n_\mu = 0$, 1,..., $N_\sigma -1$
for $\mu=1$, 2, 3 and $p_4 = (2\pi /N_\tau) n_4$, $n_4 =  0$, 1,...,
$N_\tau-1$.
The lattice gauge fields are defined on the middle of a link,
\beqn
A_\mu (p) = \sum_x {\rm e}^{\displaystyle -ipx-ip_\mu /2}
A_\mu (x+\hat\mu /2)~~,
\label{Afield}
\eqn
with $A_\mu \equiv A_\mu^a \lambda^a$, $a=1,...,~(N^2-1)$, and the
normalization $2 {\rm Tr}  \lambda^a \lambda^b = \delta_{ab}$. We also will
use the short hand notation $s_\mu \equiv \sin (p_\mu /2)$.

For the evaluation of the leading order term in the energy density on finite
lattices one only needs to keep the quadratic part in the perturbative
expansion of the action, $\beta S^I$, and introduce a gauge fixing term.
We combine both in the form
\beqn
S_0 = -{1\over 2} \int_p \sum_{\mu,\nu} A_\mu^a(-p) \Delta_{\mu, \nu} (p)
A_\nu^a(p)~~,
\label{weakaction}
\eqn
with the inverse propagator
\beqn
\Delta_{\mu, \nu} (p) = G_{\mu, \nu} (p)  +
\xi g_\mu (p) g_\nu (p)   ~~.
\label{prop}
\eqn
The first term in Eq.~(\ref{prop}) arises from
the expansion of the action and the second term defines the covariant
gauge fixing.  It is convenient to separate in $G_{\mu, \nu}$ a diagonal part,
\beqn
G_{\mu, \nu} (p) = D_\mu(p) \delta_{\mu,\nu} -
E_{\mu, \nu} (p)~~,
\label{prop11}
\eqn
with
\begin{eqnarray}
D_\mu (p) &=&  \sum_{l=1}^\infty \sum_{k=1}^l a_{k,l}\sum_{\nu=1}^4
N^{k,l}_{\mu;\nu} (p) ~~, \nonumber \\
E_{\mu,\nu} (p) &=&  \sum_{l=1}^\infty \sum_{k=1}^l a_{k,l} M^{k,l}_{\mu,
\nu} (p) ~~,
\label{pr11}
\end{eqnarray}
where $N^{k,l}_{\mu;\nu} (p)$ and $ M^{k,l}_{\mu, \nu} (p)$ are obtained
in $O(g^2)$ from the expansion of symmetrized $k\times l$ Wilson loops,
\begin{eqnarray}
W^{k,l}_{\mu, \nu} &\equiv& {1\over N_\sigma^3 N_\tau}
\sum_x W^{k,l}_{\mu, \nu} (x) \nonumber \\
&=& {g^2 \over 4N} \sum_{p, a} \biggl(
N^{k,l}_{\mu; \nu} (p) A^a_\mu (p) A^a_\mu (-p)  +
N^{k,l}_{\nu; \mu} (p) A^a_\nu (p) A^a_\nu (-p) \nonumber \\
&~&\hskip 1.0truecm  - 2 M^{k,l}_{\mu, \nu} (p)
A^a_\mu (p) A^a_\nu (-p)  \biggr) + O(g^4)~~.
\label{wexp}
\end{eqnarray}
Explicit expressions for
$N^{k,l}_{\mu;\nu} (p)$ and $ M^{k,l}_{\mu, \nu} (p)$ are given in
the Appendix.

With these definitions at hand, it is easy to evaluate the leading order
expression for the energy density on finite lattices. Using the $O(g^2)$
expansion coefficients for $k\times l$
Wilson loops, given in the Appendix, one finds,
\beqn
S^{(k, l),(2)}_{\mu,\nu} =   \int_p \biggl(
N^{k,l}_{\mu; \nu} \Delta_{\mu , \mu}^{-1}+
N^{k,l}_{\nu; \mu} \Delta_{\nu , \nu}^{-1}  -
2 M^{k,l}_{\mu, \nu} \Delta_{\mu , \nu}^{-1} \biggr)~~,
\label{waction}
\eqn
and finally
\begin{eqnarray}
S^{I,(2)}_{\sigma} &=& {1\over 3}\sum_{\mu<\nu<4} \sum_{l=1}^\infty
\sum_{k=1}^l
a_{k,l} S^{(k,l),(2)}_{\mu,\nu}
 ~~, \nonumber \\
S^{I,(2)}_{\tau} &=&   {1\over 3}\sum_{\mu=1}^3 \sum_{l=1}^\infty \sum_{k=1}^l
a_{k,l} S^{(k,l),(2)}_{\mu,4}~~.
\label{coefficient}
\end{eqnarray}

The necessary inversion of the matrix $\Delta_{\mu,\nu}$ and the evaluation
of the sums appearing in Eq.~(\ref{coefficient}) have been performed using
Mathematica. We have explicitly checked that all our results are independent
of the choice of the gauge fixing function $g_\mu (p)$ and the
gauge fixing parameter $\xi$.
In particular we have verified that in the infinite volume
limit the $O(g^2)$ expansion coefficients are identical
for all improved Wilson actions,
\beqn
\lim_{N_\tau \rightarrow \infty}
\lim_{N_\sigma \rightarrow \infty} S^{I,(2)}_{\sigma} \equiv
\lim_{N_\tau \rightarrow \infty}
\lim_{N_\sigma \rightarrow \infty} S^{I,(2)}_{\tau} \equiv
{ 1 \over 2} ~~.
\label{s2infinity}
\eqn
We also have evaluated separately the $O(g^2)$ expansion coefficients for the
expectation values of the different loops appearing in the improved actions.
For $S^{(1,2)}$ our results on an infinite lattice agree with earlier
calculations \cite{Weisz}, while we disagree for $S^{(2,2)}$ with results
quoted in \cite{Mor93}\footnote{The gluon propagator given in
Ref.\cite{Mor93} for the action $S^{(2,2)}$ corresponds to a gauge, in which
$\Delta_{\mu,\nu} = D_\mu \delta_{\mu,\nu}$. In general, this cannot be
achieved with the covariant gauge fixing term given
in Eq.~(\ref{prop}) and therefore does
require non-trivial ghost contributions already in leading order perturbation
theory, \ie~for the ideal gas \cite{Dol74}. The ghost contribution has
not been calculated in Ref.\cite{Mor93}. For this reason the $O(g^2)$ result
quoted there is not the complete contribution to the expansion
coefficient for the plaquette expectation value at that order in $g^2$.}.
For the expansion of the plaquette expectation value we find
\begin{eqnarray}
\langle 1- {1\over N} {\rm Re~Tr} U_{x,\mu}
U_{x+\hat\mu,\nu} U^{+}_{x+\hat\nu,\mu} U^{+}_{x,\nu} \rangle &=&
g^2 {N^2 - 1 \over 4N} \cases{ 0.5 &, I=(1,1) \cr
0.366263 &, I=(1,2) \cr
0.392641 &, I=(2,2) \cr
0.358304 &, I=(3,3) }
\nonumber \\
& &+ O(g^4)~~~.
\label{plaqexp}
\end{eqnarray}

\subsection{Cut-off Effects in the Wilson Action}

For the one-plaquette action we use the gauge fixing function $g_\mu =2 s_\mu$,
which, to lowest order in $a$, is the momentum representation of the
lattice version of $\partial_\mu A_\mu$.
Choosing the Feynman gauge, $\xi=1$,
we then obtain a diagonal propagator
\beqn
\Delta_{\mu, \nu} (p) = D(p) \delta_{\mu, \nu}\quad {\rm with}\quad
D(p)= 4\sum_{i=1}^4 s_i^2 ~~.
\label{diag11}
\eqn

This leads to the familiar result for the energy density
\cite{Hel85},
\beqn
{\epsilon \over T^4} = 6 (N^2 - 1) N_\tau^4 \int_p {s^2_1 - s^2_4 \over
D(p)}~~.
\label{epert}
\eqn

\begin{figure}[htb]
\centerline{
   \epsfig{bbllx=50,bblly=70,bburx=580,bbury=760,
       file=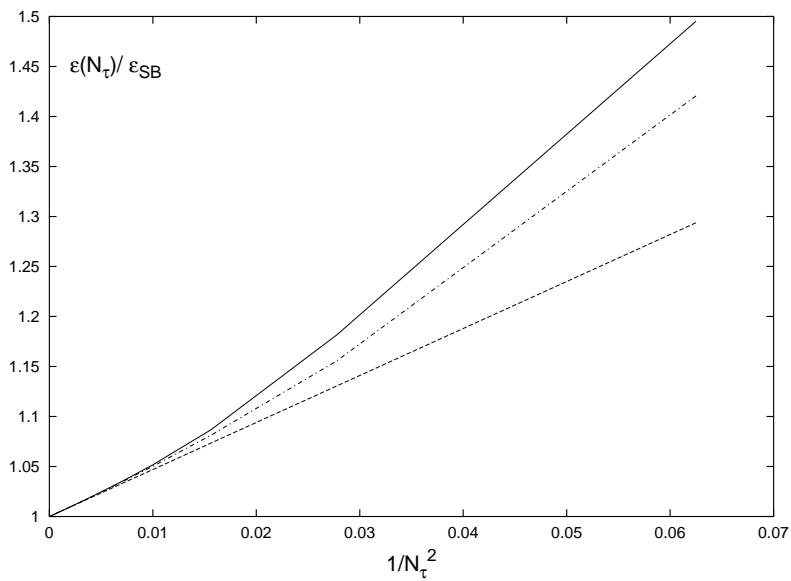,width=80mm,angle=-90}}
\caption{\baselineskip 10pt
High temperature limit of the energy density of an $SU(N)$ gluon gas
calculated with finite cut-off, $aT=1/N_\tau$, in units of the
continuum limit result (solid curve). The dashed and dashed-dotted curves
show this ratio calculated up to $O(N_\tau^{-2})$ and $O(N_\tau^{-4})$,
respectively.}

\label{fig:energy11}
\end{figure}

Eq.~(\ref{epert}) is the starting point for a discussion of the
cut-off dependence of the high temperature limit of the energy density.
We will consider here the cut-off dependence in the thermodynamic limit
($N_\sigma \rightarrow \infty$) and will comment on the influence of a finite
spatial volume later. In the thermodynamic limit the sum over the spatial
momenta can be replaced by an integral. The remaining sum over $p_4$ can be
performed explicitly \cite{Elz88}. One then finds,
\beqn
{\epsilon \over T^4} = 4  (N^2 - 1) N_\tau^4 {1 \over (2\pi)^3}
\int_0^{2\pi} {\rm d}^3 p {\omega \over \sqrt{1+ \omega^2} }
{1 \over \exp{(N_\tau x)} -1} ~~,
\label{infenergy}
\eqn
with $x =2 \ln (\omega + \sqrt{1+ \omega^2})$ and $\omega^2= s^2_1 + s^2_2
+s^2_3$.
The right hand side of the above equation can be expanded in powers of
$1/N_\tau \equiv aT$, which explicitly shows that the leading corrections to
the continuum Stephan-Boltzmann law are indeed $O((aT)^2)$ \cite{Eng95}.
This expansion is somewhat tedious but straightforward. In next-to-leading
order we find\footnote{Note that the natural expansion parameter
is $\pi Ta \equiv \pi/N_\tau$, \ie~half the smallest, non-vanishing
Matsubara mode.}
\beqn
{\epsilon \over T^4} = 3{p \over T^4} =
(N^2-1) {\pi^2 \over 15}\biggl[ 1+
{30  \over 63} \cdot { \pi^2 \over N_\tau^2} +
{1 \over 3} {\pi^4 \over N_\tau^4} +
O\biggl ({1 \over N_\tau^6} \biggr ) \biggr ]~.
\label{Ntdependence}
\eqn
In Figure~\ref{fig:energy11} we show the cut-off dependence of
$\epsilon / T^4$ obtained from Eq.~(\ref{infenergy}) as well as
the $O(N_\tau^{-2})$ and $O(N_\tau^{-4})$ results from
Eq.~(\ref{Ntdependence}).
As can bee seen the leading finite size correction term given in
Eq.~(\ref{Ntdependence}) accounts for more than 70\% of the total
correction already for $N_\tau \ge 6$.

\subsection{Improved Actions}

Generalizing the discussion of the perturbative calculations with the
one-plaquette Wilson action to the case of the improved actions is now
straightforward, although the actual calculations turn out to be quite
tedious because we did not succeed in finding a gauge fixing condition,
which would lead to a simple diagonal propagator. Details of the perturbative
calculations for the various improved actions considered by us are given
in the Appendix.

The generalization of the leading order perturbative expansion for the
energy density, given in Eq.~(\ref{epert}) for the one-plaquette Wilson
action, now reads
\beqn
{\epsilon \over T^4} = {3 \over 2} (N^2-1) N_\tau^4 \int_p
{R_\sigma (p) - R_\tau (p) \over R^d (p)} ~~,
\label{improvedenergy}
\eqn
with functions $R_\sigma,~R_\tau$ and $R^d$ defined in the Appendix for
the different improved actions.

We have evaluated the energy density from Eq.~(\ref{improvedenergy})
numerically.
Results for small values of $N_\tau$ obtained in the thermodynamic limit
($N_\sigma \rightarrow \infty$) are summarized in
Table~\ref{tab:improvement}. We note that for all improved actions the
deviation from the continuum results, $\epsilon_{\rm SB}/T^4$, is less
than $1.2$\% already for $N_\tau =6$ which is reached with the Wilson action
only for $N_\tau \simeq 20$. At this level of accuracy
it is, in fact, important to consider also the influence of a
finite spatial extent of the lattice. A calculation of $\epsilon (N_\tau,
N_\sigma) / \epsilon_{\rm SB}$ for finite $N_\sigma$ indeed shows that
the resulting infrared finite-size effects also are of the order of 1\% for
$N_\sigma / N_\tau \simeq 4$. Some results for finite spatial lattices are
summarized in Table~\ref{tab:finitesize}.

\begin{table}[hbt]
\newlength{\digitwidth} \settowidth{\digitwidth}{\rm 0}
\catcode`?=\active \def?{\kern\digitwidth}
\caption{Deviations from the continuum ideal gas behaviour on spatially
infinite lattices with temporal extent $N_\tau$ for the standard one-plaquette
Wilson action and various improved actions. In the last row we give
the coefficients of the leading cut-off correction term as defined in
Eq.~(3.19).
The number in brackets gives the difference between the estimate for $c^I$
obtained for $N_\tau=32$ and the extrapolated value.}

\label{tab:improvement}
\vskip 5pt
\begin{tabular*}{\textwidth}{@{}l@{\extracolsep{\fill}}ccccc}
\hline
        & \multispan4  $\epsilon (N_\tau) / \epsilon_{\rm SB}$ \\
\hline
$N_\tau$ & I=(1,1)& I=(1,2) & I=(2,2)& I=(3,3) \\
\hline
{}~4 & 1.495186 & 0.986568  & 1.087709 & 1.044357 \\
{}~6 & 1.181566 & 0.997528  & 1.011994 & 1.008095 \\
{}~8 & 1.086700 & 1.000309  & 1.003752 & 1.001199 \\
10 &1.051708 & 1.000253  & 1.001582 & 1.000320 \\
12 &1.034756 & 1.000150  & 1.000780 & 1.000110 \\
\hline

\hline
$c^I$ & 0.4762 & 0.044~(2)   & 0.178~(2)  & 0.394~(19) \\
\hline
\end{tabular*}

\end{table}

\begin{table}[hbt]
\catcode`?=\active \def?{\kern\digitwidth}
\caption{Deviations from the continuum ideal gas behaviour on finite
spatial lattices with temporal extent $N_\tau$ for the standard one-plaquette
Wilson action and various improved actions. }

\label{tab:finitesize}
\vskip 5pt
\begin{tabular*}{\textwidth}{@{}l@{\extracolsep{\fill}}|cccc|cccc}
\hline
        & \multispan8  $\epsilon (N_\tau,N_\sigma) / \epsilon_{\rm SB} $ \\
\hline
     &  & \multispan2  $N_\sigma =4 N_\tau$ & &
        & \multispan2  $N_\sigma =6 N_\tau$ & \\
\hline
$N_\tau$ & I=(1,1)& I=(1,2) & I=(2,2)& I=(3,3) & I=(1,1) & I=(1,2) & I=(2,2)
& I=(3,3)\\
\hline
{}~4 & 1.4833 & 0.9747 & 1.0758  & 1.0325 & 1.4917 & 0.9830 & 1.0842 & 1.0408
\\
{}~6 & 1.1697 & 0.9857 & 1.0001  & 0.9962 & 1.1780 & 0.9940 & 1.0085 & 1.0046
\\
{}~8 & 1.0748 & 0.9884 & 0.9919  & 0.9893 & 1.0832 & 0.9968 & 1.0002 & 0.9977
\\
10 & 1.0398 & 0.9884 & 0.9897  & 0.9884 & 1.0482 & 0.9967 & 0.9981 & 0.9968 \\
12 & 1.0229 & 0.9883 & 0.9889  & 0.9882 & 1.0312 & 0.9966 & 0.9973 & 0.9966 \\
\hline

\end{tabular*}
\end{table}

\begin{figure}[htb]
\centerline{
   \epsfig{bbllx=70,bblly=70,bburx=565,bbury=765,
       file=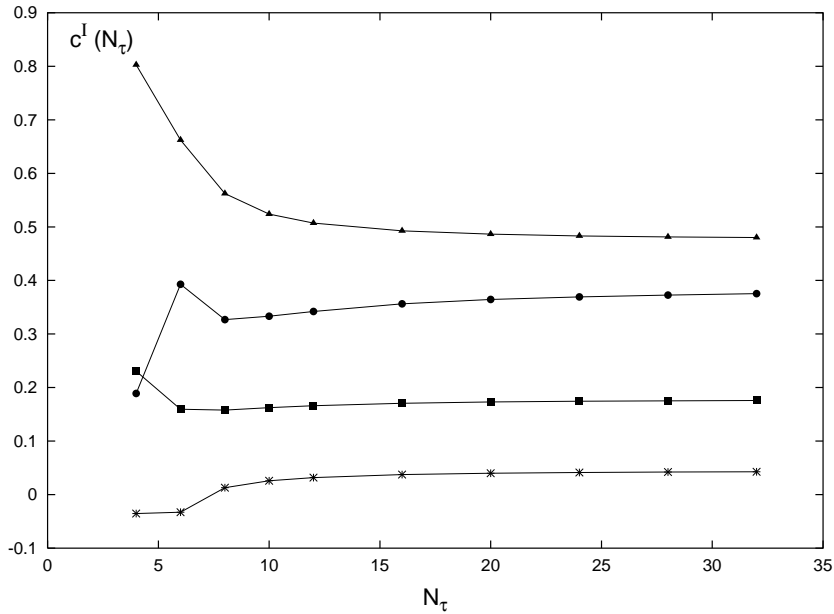,width=80mm,angle=-90}}
\caption{\baselineskip 10pt
Estimate of the coefficient $c^I$ of the leading cut-off dependence, obtained
by solving Eq.~(3.19) for $c^I$ at fixed $N_\tau$, with $\alpha = 2$ for the
one-plaquette Wilson action (triangle), $\alpha = 4$ for  $I=(1,2)$
(crosses), $I=(2,2)$ (circle) and $\alpha = 6$ for $I=(3,3)$
(squares).  In the limit $N_\tau \rightarrow \infty$ this quantity
yields the coefficient $c_I$ given in Table 1.}

\label{fig:eimproved}
\end{figure}

The energy density calculated perturbatively from the improved actions
will show a cut-off dependence starting at $O((aT)^\alpha)$, \ie~corrections
on lattices with temporal extent $N_\tau$ will asymptotically be proportional
to $N_\tau^{-\alpha}$ with $\alpha = 4$ for the cases $I=(1,2)$ an $I=(2,2)$
and  $\alpha = 6$ for $I=(3,3)$,
\beqn
{\epsilon (N_\tau) \over T^4} = (N^2-1){\pi^2 \over 15} \biggl[ 1 +
c^I \biggl({ \pi \over N_\tau}\biggr)^\alpha + O(N_\tau^{-(\alpha+2)})
\biggr] ~~.
\label{calpha}
\eqn
We have evaluated $\epsilon/T^4$ also for large values of $N_\tau$,
up to $N_\tau \le 32$, in order to verify this asymptotic behaviour. In
Figure~\ref{fig:eimproved} we show estimates for the expansion
coefficients, $c^I(N_\tau)$, obtained at fixed values of $N_\tau$ by solving
Eq.~(\ref{calpha}) for $c^I$. As can be seen,
the cut-off dependence indeed shows the anticipated scaling behaviour.
The coefficients $c^I$ have then been determined from an extrapolation
to $N_\tau \rightarrow \infty$, taking into account the subleading correction
term proportional to $N_\tau^{-(\alpha+2)}$. These numbers are given in
the last row of  Table~\ref{tab:improvement}. The quoted errors reflect
the difference between the extrapolated value and the last calculated
approximant for $N_\tau=32$. We note that in the case of the improved actions
$S^{(1,2)}$ and $S^{(2,2)}$ not only the $O(N_{\tau}^{-2})$ corrections have
been eliminated, but also the magnitude of the $O(N_{\tau}^{-4})$ coefficient
has been reduced strongly when compared to the one-plaquette action.

The improved actions lead
to a drastic reduction of finite cut-off effects already for small values
of $N_\tau$, where in the case of the one-plaquette Wilson actions the
cut-off effects were not at all dominated by the leading $O(N_\tau^{-2})$
contribution, which has been eliminated in the improved actions. In fact,
in the case of the improved action $S^{(1,2)}$, deviations from the continuum
ideal gas behaviour are always less than $1.5\%$ already for $N_\tau \ge4$,
which is
to be compared with the nearly 50\% deviations at $N_\tau =4$ for the
one-plaquette Wilson action. We note, however, that also in this case the
leading $N_\tau^{-4}$
dependence of the cut-off effects starts dominating only for $N_\tau \ge 10$.
This is different for the actions $S^{(2,2)}$ and $S^{(3,3)}$, where the
leading correction dominates already for $N_\tau \ge 6$.

\section{Monte Carlo Results for Improved Actions}

The perturbative calculations have shown that already
the $O(a^2)$ improved actions $I=(2,2)$ and in particular $I=(1,2)$
provide a large reduction of finite cut-off effects in the high
temperature ($T\rightarrow \infty$) limit even for small values of
$N_\tau$. We have used these actions as well as the $O(a^4)$ improved action,
$I=(3,3)$, to analyze the cut-off dependence at
non-vanishing values of $g^2$, \ie~ for $T<\infty$. The recently performed
simulations with
the one-plaquette Wilson action on lattices with $N_\tau =4$, 6 and 8
\cite{Boyd95} do provide here a good basis for a comparison.

In a first step we have evaluated the difference of action densities,
studied perturbatively in the previous sections, at finite values of the
gauge coupling $\beta$ on lattices of size $16^3\times 4$ ($24^3 \times 4$
in the case of the action $S^{(2,2)}$). We have
performed calculations of the action differences for the $SU(3)$ gauge
theory,
\beqn
\biggl({\epsilon \over T^4}\biggr)_0 = 3 \beta N_\tau^4 \biggl( \langle
S^{I}_{\sigma}\rangle -  \langle S^{I}_{\tau}\rangle \biggr)~~,
\label{epsilon_0}
\eqn
at values of $\beta=6/g^2$ well above the critical value for the
deconfinement transition. Some results are given in Table~\ref{tab:actiondif}.
As can be seen, also at finite values of $\beta$ the results for the
different actions show a cut-off dependence, which is close to the
result calculated in the limit $\beta \rightarrow \infty$. In all cases the
asymptotic value is approached from above. For the
one-plaquette Wilson action this has been shown to be in accordance
with the perturbatively calculated $O(g^2)$ correction to
$ ({\epsilon / T^4})_0 $ \cite{Hel85}. This behaviour should not be confused
with
the correction to the complete energy density, $ {\epsilon / T^4}$, which
is negative.

\begin{table}[hbt]
\catcode`?=\active \def?{\kern\digitwidth}
\caption{Differences of space- and timelike action densities times $N_\tau^4$
calculated on lattices of size $16^3\times 4$ for $I=(1,1)$, (1,2) and (3,3)
and $24^3\times 4$ in the case of $I=(2,2)$. The last row gives the
perturbative result in the limit $\beta \rightarrow \infty$ (the corresponding
value in the continuum is $ \epsilon_{\rm SB} / T^4 = 5.2638$). Couplings have
been chosen well above the critical coupling for the deconfinement
transition, which occurs at $\beta_c = 5.6908~(2)$ ($I=(1,1)$, [20]),
$  4.0752~(13)$ ($I=(1,2)$, [7]),
$  4.3995~(2)$ ($I=(2,2)$, [this study]) and
$  3.7 < \beta_c < 4.2$ ($I=(3,3)$, [this study]).}

\label{tab:actiondif}
\vskip 5pt
\begin{tabular*}{\textwidth}{@{}l@{\extracolsep{\fill}}ccccc}
\hline
&\multispan4 $ ({\epsilon / T^4})_0 $ \\
\hline
$\beta $ & (1,1) & (1,2) & (2,2) & (3,3) \\
\hline
{}~6.0 & 7.359~(~12) & 5.706(57)  & 6.513~(74) & 5.88~(12) \\
10.0 & 8.046~(102) & 5.42~(10)  & 6.203~(90) & 5.79~(16) \\
15.0 & 7.837~(~63) & 5.445(84)  & 5.982~(79) & 5.58~(16) \\
20.0 & 7.806~(~48) & 5.18~(10)  & 5.860~(94) & 5.56~(13) \\
\hline
$\infty$ & 7.808 & 5.131  & 5.707 & 5.435 \\
\hline
\end{tabular*}

\end{table}

A more detailed comparison with the one-plaquette Wilson action is possible
through a complete calculation of the temperature dependence of bulk
thermodynamic quantities along the line presented in \cite{Boyd95}. This
also requires the calculation of the action densities on large, zero
temperature lattices. As this is quite time consuming we have performed
such a complete analysis only for the action $I=(2,2)$. On a lattice with
four sites in the temporal direction this action leads,
in the limit $T \rightarrow \infty$, to finite cut-off
effects which are of similar size as those found
for the one-plaquette Wilson action on lattices
with eight sites in the temporal direction. A direct comparison of results
obtained with both action thus is possible.

We have calculated
the action densities $\langle S^{(2,2)} \rangle_T$ and
$\langle S^{(2,2)} \rangle_0$ on lattices of size $24^3\times 4$
and $24^4$, respectively. An integration of their differences calculated
at various values of $\beta$ yields the pressure\footnote{We refer to
Refs.~\cite{Eng95, Boyd95} for more details on the formalism.}
\beqn
{p\over T^4}\Big\vert_{\beta_0}^{\beta}
=~N_\tau^4\int_{\beta_0}^{\beta}
{\rm d}\beta'  (\langle S^{(2,2)} \rangle_0 -
\langle S^{(2,2)} \rangle_T )~~,
\label{freelat}
\eqn
where $\beta_0$ is a value of the coupling constant below the phase transition
point at which the pressure can savely be approximated by zero.
In order to compare this calculation with corresponding results for the
one-plaquette Wilson action, we have to define a common temperature scale.
For this purpose we determine the relation between $\beta$ and the cut-off
via the 2-loop
renormalization group equation, $a\Lambda = R(\beta_{\rm eff})$ using there
an effective coupling constant obtained in terms of the action density
\beqn
\beta_{\rm eff} = {N^2-1 \over 4 \langle S^I \rangle_0}~~.
\label{beff}
\eqn
In the case of the one-plaquette Wilson action, this has been found to
be a reasonable parametrization of the temperature scale
\cite{Boyd95}.
Combined with a determination of the critical coupling for the deconfinement
transition on the $24^3\times 4$ lattice,
\beqn
\beta_c (N_\tau=4) = 4.3995 \pm 0.0002 ~~,
\label{bc}
\eqn
the temperature can be expressed in units of the critical
temperature $T/T_c = R(\beta_{{\rm eff},c}) /  R(\beta_{\rm eff})$.

\begin{figure}[htb]
\centerline{
   \epsfig{bbllx=60,bblly=70,bburx=570,bbury=760,
       file=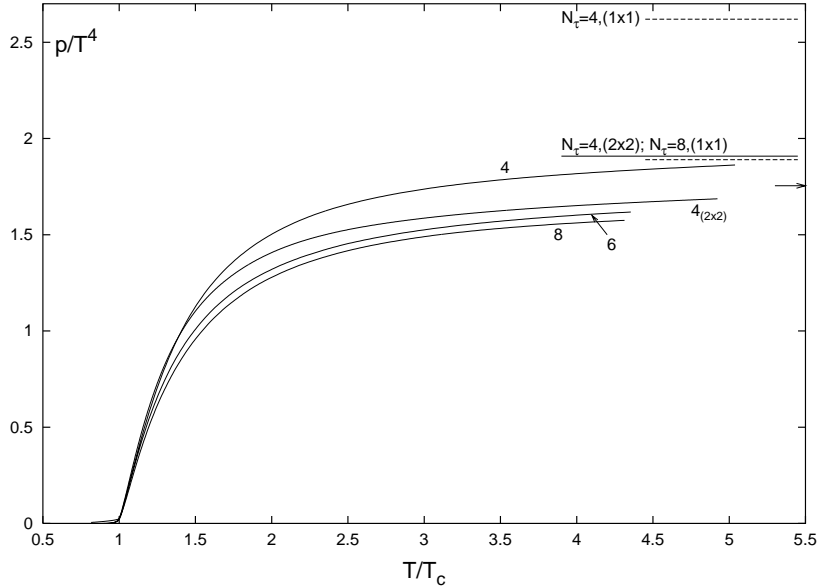,width=80mm,angle=-90}}
\caption{\baselineskip 10pt
The pressure in units of $T^4$ versus $T/T_c$ calculated with the improved
action $I=(2,2)$ on a $24^3\times 4$ lattice. This is compared
with calculations using the standard one-plaquette Wilson action on lattices
of size $32^3\times 4$, 6 and 8 [14]. The dashed horizontal lines give the
corresponding ideal gas limits for the $(1\times1)$ Wilson action and the
solid line corresponds to the ideal gas limit for the $(2,2)$-improved
action (see Table 2). The arrow points at the continuum result for an ideal
gas.}
\label{fig:pressure}
\end{figure}

In Figure~\ref{fig:pressure} we show the pressure calculated with the action
$S^{(2,2)}$ on a $24^3\times 4$ lattice and compare this with the
calculations performed with the one-plaquette Wilson action on lattices
of size $32^3\times 4$, 6 and 8 \cite{Boyd95}.

We note that at high temperatures, $T \gsim 4T_c$,
the pressure calculated from the improved action, $S^{(2,2)}$,
on a lattice with $N_\tau =4$ indeed is in good agreement with results
obtained for the one-plaquette Wilson action on a lattice with $N_\tau =8$.
This confirms that the improvement of the actions at $g^2=0$ persists
also at non-vanishing values of the gauge coupling.
For temperatures closer to $T_c$ we find, however, results, which are
compatible with those obtained with the one plaquette Wilson action.
This is not too surprising. Already the analysis of the cut-off effects in
calculations with the Wilson action has shown that the magnitude of the
cut-off corrections does vary with temperature.

\section{Conclusions}

In the high temperature limit the thermodynamics of $SU(N)$ gauge theories
is dominated by high momentum modes. In lattice calculations these modes
are most strongly affected by the finite ultraviolet cut-off. We have
shown here that the use of tree level improved actions for the gauge
fields indeed leads to a strong reduction of finite cut-off effects in
thermodynamic observables like the energy density or the pressure. The
perturbative analysis of the high temperature limit of the energy density
has shown that already on lattices with a temporal extent as small as
$N_\tau=4$ the cut-off effects can easily be reduced to a few percent.

A first Monte Carlo calculation of the pressure of an $SU(3)$ gauge theory
using an improved action has shown that this also removes the dominant
cut-off effects at non-zero values of the gauge coupling. At high
temperatures the calculations
performed with an improved action on a lattice with $N_\tau=4$
lead to results which are compatible with those obtained with the standard
one-plaquette Wilson action on a lattice with $N_\tau=8$, where this action
leads to similarly small cut-off effects. However, closer to $T_c$ we also
find that the inherently non-perturbative features of the deconfinement
phase transition play a much more important role. Here the results obtained
with a tree level improved action coincide with those obtained with the
standard Wilson action on the same size lattice. Clearly in this temperature
region there is need for further improvement of the lattice regularized
Yang-Mills action. This can, for instance, be achieved by including one-loop
corrections in the coefficients $a_{k,l}$ of the improved action or by
determining them through a Monte Carlo renormalization group approach at
various values of $g^2$. Currently we also investigate the properties of
tadpole improved actions \cite{Alf95}.

Of course, considerations similar to those made here for $SU(N)$ gauge
theories can also be applied to the fermion sector of QCD. In fact, in
this case the influence of finite cut-off effects on the high temperature
limit of thermodynamic observables are known to be even larger on lattices
with small temporal extent. The calculation of the equation of state of QCD
thus should profit even more from the use of improved actions.
An investigation of this claim is under way.

\vskip 10pt
\noindent
{\large \bf Acknowledgements:}
\vskip  5pt
\noindent
The numerical calculations presented here have been performed on a
QUADRICS parallel computer funded by the
Deutsche Forschungs\-gemeinschaft (DFG) under grant Pe 340/6-1.
It also has been supported by the DFG through the grants Pe 340/3-2 and
Pe 340/3-3. F.K. thanks the theory division of CERN for the kind
hospitality extended to him during the summer 1995.

\appendix
\section{Appendix}

We give here explicit expressions for the leading order weak coupling
expansion of the action densities on
finite lattices for the improved Wilson actions defined in
Eqs.~(\ref{actionnn} - \ref{actions}), respectively. For completeness we also
repeat the corresponding results for the standard one-plaquette Wilson action.
In all cases we have checked that the results are gauge invariant.

In addition to the short hand notation $s_\mu \equiv \sin(p_\mu /2)$
we introduce the abbreviations
\begin{eqnarray}
\Delta^{-}_\mu (p) &=& e^{i p_\mu} - 1 \nonumber ~~,\\
\Delta^{(k)}_\mu (p) &=& \sum_{l=0}^{k-1} e^{ilp_\mu} ~~.
\label{dp}
\end{eqnarray}
The expansion of a symmetrized $k\times l$ Wilson loop defined in
Eq.~(\ref{wloop}) to $O(g^2)$ can then be written as
\begin{eqnarray}
W^{k,l}_{\mu, \nu} &\equiv& {1\over N_\sigma^3 N_\tau}
\sum_x W^{k,l}_{\mu, \nu} (x) \nonumber \\
&=& {g^2 \over 4N} \sum_{p, a} \biggl(
N^{k,l}_{\mu; \nu} (p) A^a_\mu (p) A^a_\mu (-p)  +
N^{k,l}_{\nu; \mu} (p) A^a_\nu (p) A^a_\nu (-p) \nonumber \\
&~&\hskip 1.0truecm  - 2 M^{k,l}_{\mu, \nu} (p)
A^a_\mu (p) A^a_\nu (-p)  \biggr)  +O(g^4)
\label{wexpb}
\end{eqnarray}
with
\beqn
N^{k,l}_{\mu; \nu} (p) =  {1 \over 2} \biggl( \Delta^{(k)}_\mu (p)
\Delta^{(k)}_\mu (-p)  \Delta^-_\nu (-lp) \Delta^-_\nu (lp) +
({k \leftrightarrow l})\biggr)
\label{nfourier}
\eqn
and
\beqn
M^{k,l}_{\mu, \nu} (p) =  {1 \over 2} \biggl(
e^{i (p_\mu-p_\nu) /2} \Delta^{(k)}_\mu (p)
\Delta^-_\nu (lp) \Delta^{(l)}_\nu (-p)
\Delta^-_\mu (-kp) + ( k \leftrightarrow l) \biggr) ~~.
\label{wfourier}
\eqn

We note that $M^{k,l}_{\mu,\nu}$ has non-vanishing matrix
elements also for $(\mu = \nu)$. In fact, due to the identity
$\Delta^{(k)}_\mu (p) \Delta^-_\mu (lp) =
\Delta^{(l)}_\mu (p) \Delta^-_\mu (kp)$
one finds that $M^{k,l}_{\mu,\mu} = N^{k,l}_{\mu;\mu} $.
This term does not contribute in the expansion of Wilson loops. It is
included in the definition of $E_{\mu,\nu}$ and
cancels against the corresponding term in the
diagonal part of the matrix $G_{\mu,\nu} = D_\mu(p) \delta_{\mu,\nu} -
E_{\mu, \nu} (p)$, which has been introduced in  Eq.~(\ref{pr11}).
With the above relations we obtain for $ D_\mu(p)$ and $E_{\mu, \nu}
(p)$,
\begin{eqnarray}
D_\mu (p) &=&  \sum_{l=1}^\infty \sum_{k=1}^l a_{k,l} \sum_{\nu=1}^4
N^{k,l}_{\mu; \nu} (p)\nonumber \\
&=&  2 \sum_{l=1}^n \sum_{k=1}^l a_{k,l} \biggl(
\biggl[k+2\sum_{j=1}^{k-1} (k-j)\cos(jp_\mu)\biggr] \sum_{\nu=1}^4
\sin^2(l p_\nu /2)
\nonumber \\
& &\qquad\qquad +   ({k \leftrightarrow l})\biggr)
\nonumber \\
E_{\mu,\nu} (p) &=&  \sum_{l=1}^n \sum_{k=1}^l a_{k,l} M^{k,l}_{\mu,
\nu} (p)
\\
&=& 2 \sum_{l=1}^n \sum_{k=1}^l a_{k,l} \biggl(
\sum_{i=0}^{k-1} \sin((2i+1)p_\mu/2) \sum_{j=0}^{l-1}
\sin((2j+1)p_\nu/2)
\nonumber \\
& &\qquad\qquad +   ({k \leftrightarrow l})\biggr)
\nonumber
\label{eqXY}
\end{eqnarray}

Adding the gauge fixing functions $g_\mu$
one can construct the propagator matrix $\Delta_{\mu,\nu}$ (Eq.~(\ref{prop})),
which we invert using Mathematica.

The explicit results for the standard one-plaquette Wilson action have been
discussed in Section 3.1. We give here the result for the expansion
coefficient, $S^{I,(2)}_{\mu,\nu}$, for improved actions, which we write in the
form
\beqn
S^{I,(2)}_{\mu,\nu} = \int_p {R_{\mu,\nu} (p) \over
R^d (p)}~~.
\label{scoef}
\eqn
In Eq.~(\ref{improvedenergy}) we have introduced the functions
$R_\sigma$ and $R_\tau$, which are defined as
\begin{eqnarray}
R_\sigma &=& {1\over 3} (R_{1,2} + R_{1,3} + R_{2,3} ) ~~,
\nonumber \\
R_\tau &=& {1\over 3} (R_{1,4} + R_{2,4} + R_{3,4} ) ~~.
\label{rsrt}
\end{eqnarray}

In order to present the results in a convenient form, we have chosen
a specific form of the gauge fixing function
\begin{eqnarray}
I=(1,1) &:& g_\mu =  2\; \sm  \nonumber \\
I=(1,2) &:& g_\mu =   2\; ( \sm + \frac{2}{3}~\smb )  \nonumber \\
I=(2,2) &:& g_\mu =  2\; ( \sm + \smb )  \nonumber \\
I=(3,3) &:& g_\mu =  2\; ( \sm +  \frac{1}{3} \smb + \frac{16}{45} \smc )~~.
\label{gaugef}
\end{eqnarray}

In the following we use the convention that all indices specified are
mutually different. Permutations of indices that lead to identical
expressions are counted only once. Whenever summations over permutations of
indices are needed the relevant sums are specified explicitly. We present the
result for the $(\mu, \nu)$ component of the action. The remaining two
directions of the 4-dimensional lattice are denoted by $(i,j)$. In those cases
where indices can take values from the set  $(\mu, \nu)$ as well as  $(i,j)$ we
denote them by $k,~l,~m$ and $n$.  With these conventions we find
\vskip 10pt
\noindent
$I=(1,1):$
\begin{eqnarray}
R_{\mu,\nu} &=& s_\mu^2+s_\nu^2
\nonumber \\
R^d &=&   \sum_{k=1}^4 s_k^2
\label{s11}
\end{eqnarray}
\vfill\eject
\noindent
$I=(1,2):$
\begin{eqnarray}
R_{\mu,\nu} &=& \frac{4}{3} (3+s_\mu^2+s_\nu^2)
\nonumber \\
& &\quad \biggl( D_i D_j (s_\mu g_\mu+ s_\nu g_\nu)^2
+(D_\mu s_\mu^2  + D_\nu s_\nu^2 )(D_i g_j^2  + D_j g_i^2 ) \biggr)
\nonumber \\
R^d &=& \sum_{k=1}^4 g_k^2 D_l D_m D_n
\label{s12}
\end{eqnarray}
\vskip 10pt
\noindent
$I=(2,2):$
\begin{eqnarray}
R_{\mu,\nu} &=& {4 \over 3} (3 + s_\mu^2+s_\nu^2 - s_\mu^2 s_\nu^2 )
\nonumber \\
& &\quad \biggl(D_i D_j (s_\mu g_\mu+s_\nu  g_\nu)^2
+(D_\mu s_\mu^2 + D_\nu s_\nu^2 ) ( D_i g_j^2 +D_j g_i^2) \biggr)
\nonumber \\
R^d &=& \sum_{k=1}^4 g_k^2  D_l D_m D_n
\label{s22}
\end{eqnarray}
\noindent
$I=(3,3):$
\begin{eqnarray}
R_{\mu,\nu} &=& 4 \biggl( 1 +{1\over 3} (s_\mu^2+s_\nu^2) +
{8 \over 45} (s_\mu^4 +s_\nu^4)
+{1\over 9} s_\mu^2 s_\nu^2 - {64\over 135} s_\mu^2 s_\nu^2 (s_\mu^2 +s_\nu^2)
+ {128\over 405} s_\mu^4 s_\nu^4 \biggr)
\nonumber \\
& &\biggl[ D_i D_j (s_\mu g_\mu + s_\nu g_\nu )^2
+ (D_\mu s_\mu^2 + D_\nu s_\nu^2) (D_i g_j^2 +D_j g_i^2)
\nonumber \\
& &+{256 \over 18225} (D_\mu s_\mu^2 + D_\nu s_\nu^2) A_{i,j}B_{j,i}
\nonumber \\
& &-{256 \over 18225} \biggl( D_i s_j^2 (A_{\mu,j} +A_{\nu,j})
(B_{\mu,j} +B_{\nu,j}) +
D_j s_i^2 (A_{\mu,i} +A_{\nu,i}) (B_{\mu,i} +B_{\nu,i}) \biggr)
\nonumber \\
& &-{4096 \over 225}  s_i^2 s_j^2(s_j^2 -s_i^2)^2 \biggl(
s_\mu^2 (s_\mu^2-s_i^2) (s_\mu^2-s_j^2) +
s_\nu^2 (s_\nu^2-s_i^2) (s_\nu^2-s_j^2) \biggr)^2~ \biggr]
\nonumber \\
& &\nonumber \\
R^d &=&
\sum_{k=1}^4 g_k^2 D_l D_m D_n
+ {256 \over 18225} \sum_{k=1}^3 \sum_{l=k+1}^4 D_k D_l A_{n,m}B_{m,n}
\nonumber \\
& &-{4096 \over 225} \sum_{k=1}^4 D_k s_l^2 s_m^2 s_n^2
(s_m^2-s_l^2)^2 (s_m^2-s_n^2)^2 (s_l^2-s_n^2)^2
\nonumber \\
\label{s33}
\end{eqnarray}
Here we also have used the abbreviations
\begin{eqnarray}
A_{k,l} &=& s_k^2 (s_k^2-s_l^2) ( s_k^2 s_l^2 + 3( s_k^2 +s_l^2) )
\nonumber \\
B_{k,l} &=& s_k^2(s_k^2-s_l^2)(-405 + 120( s_k^2 + s_l^2) + 184 s_k^2 s_l^2 )
\label{abr33}
\end{eqnarray}

\vfill\eject

\end{document}